\documentclass[showpacs,prl,10pt,a4paper,twocolumn]{revtex4}

\usepackage{times}
\usepackage{amssymb}
\usepackage{amsmath}
\usepackage[dvips]{graphicx}
\usepackage{subfigure}
\usepackage{hyperref}

\begin{document}

\title{Quantum teleportation in superconducting charge qubits}
\author{R. M. Gomes}
\affiliation{Instituto de F\'{\i}sica, Universidade Federal de Goi\'{a}s, 74.001-970, Goi%
\^{a}nia (GO), Brazil}
\author{W. B. Cardoso}
\affiliation{Instituto de F\'{\i}sica, Universidade Federal de Goi\'{a}s, 74.001-970, Goi%
\^{a}nia (GO), Brazil}
\author{A. T. Avelar}
\email{ardiley@if.ufg.br}
\affiliation{Instituto de F\'{\i}sica, Universidade Federal de Goi\'{a}s, 74.001-970, Goi%
\^{a}nia (GO), Brazil}
\author{B. Baseia}
\affiliation{Instituto de F\'{\i}sica, Universidade Federal de Goi\'{a}s, 74.001-970, Goi%
\^{a}nia (GO), Brazil}

\begin{abstract}
In this paper we propose a scheme to implement a quantum teleportation based
on the current experimental design [Nature (London) \textbf{431}, \textbf{162%
} (2004); \textit{ibid} \textbf{445}, 515 (2007)] in which superconducting
charge qubits are capacitively coupled to a single high-Q superconducting
coplanar resonator. As advantage of this architecture, it permits the use of
multiqubit gates between non-nearest qubits and the realization of parallel
gates. We consider the case of two qubits inside the resonator, where the
teleportation is accomplished.
\end{abstract}

\maketitle

Josephson junctions composing superconducting circuits are currently
considered the most experimentally advanced solid state qubits \cite%
{Wendin06}. The quantum behavior of these circuits have been tested at the
level of a single qubit \cite%
{WallraffPRL95,ChiorescuSC03,VionSC02,MartinisPRL02,NakamuraNAT99} and also
for a pair of qubits \cite%
{McDermottSC05,MajerPRL05,YamamotoNAT03,BerkleySC03,PashkinNAT03}. The first
quantitative experimental study of an entangled pair of coupled
superconducting qubits was recently reported \cite{SteffenSC06}. One of the
possible applications of them concerns the quantum memory for
superconducting qubits \cite{EmilyPRA05}. This system has great interest in
fundamentals of quantum mechanics, quantum states engineering, and has also
been proposed as a candidate for use in quantum information processing \cite%
{MabuchiSCI02,HoodSCI00,RaimondRMP01}, quantum cryptography \cite{EkertPRL91}%
, quantum computer \cite{DeutschPRSLA92}, and quantum teleportation \cite%
{BennettPRL93}.

For the implementation of quantum computer and teleportation the main
ingredient is the conditional quantum dynamics, in which the coherent
evolution of a subsystem depends on the state of another one and a
measurement made upon one of them not only gives information about the
other, but also provides its manipulation. Schemes in cavity-QED\ have also
been proposed for realizing quantum logic gates \cite{BarencoPRL95} and
teleportation \cite{DavidovichPRA94}. Quantum logic gates have been
demonstrated in cavity QED \cite{TurchettePRL95}, ion trap \cite{MonroePRL95}%
, and NMR \cite{GershenfeldSC97} experiments. The ion trap is also a good
system for quantum information processing \cite{CiracPRL95}. On the other
hand, quantum teleportation has been demonstrated using optical systems \cite%
{BouwmeesterNAT97} and NMR \cite{McDermottSC05}. In our case we will use the
circuit quantum electrodynamics architecture \cite{BlaisPRA07}, in which two
superconducting charge qubits, the Cooper pair boxes (CPBs), are strongly\
coupled to a coplanar transmission line resonator.

The description of a typical Cooper pair box (CPB) is detailled in \cite%
{WallraffNAT04} ; accordingly, it consists of a several micrometre long and
submicrometre wide superconducting island which is coupled via two
submicrometre size Josephson tunnel junctions to a much larger
superconducting reservoir, constructed in the gap between the centre
conductor and the ground plane of the resonator, at an antinode of the field
\cite{WallraffNAT04}. The overall energy scales of these terms, the charging
energy $E_{C}$ and the Josephson energy $E_{J,max}$, are readily engineered
in fabrication by the choice of the total box capacitance and resistance
respectively, subsequently tuned by electrical means. A gate voltage $V_{g}$
applied to the input port (see \cite{WallraffNAT04}) induces a gate charge $%
n_{g}=V_{g}C_{g}^{\ast }/e$ that controls the electrostatic energy $E_{el}$,
where $C_{g}^{\ast }$ is the effective capacitance between the input port of
the resonator and the island of the CPB. A flux bias $\Phi _{b}=\Phi /\Phi
_{0}$ controls $E_{J}$ via the application, with an external coil, to the
loop of the box. When concerned with the case of two qubits they are usually
fabricated at the two ends of the resonator, sufficiently far apart that the
direct qubit-qubit capacitance is negligible \cite{GywatPRB06}. An advantage
of placing the qubits at the ends of the resonator is the finite capacitive
coupling between each qubit and the input or output port of the resonator.
This can be used to independently \textit{dc} bias the qubits at their
charge degeneracy point. The size of the direct capacitance must be chosen
in such a way as to limit energy relaxation and dephasing due to noise at
the input-output ports. Some of the noise is however still filtered by the
high-Q resonator \cite{BlaisPRA04}. We note that recent design advances have
also raised the possibility of eliminating the need for \textit{dc} bias
altogether \cite{SchusterNAT07}.

In this paper we propose a scheme to implement a quantum teleportation in
which two superconducting charge qubits are capacitively coupled to a single
high-Q superconducting coplanar resonator. The qubits in the circuit QED
architecture are constituted by split-junction Cooper pair boxes \cite%
{WallraffPRL95}. These devices can be modeled as two-level systems with the
Hamiltonian \cite{MakhlinRMP01}
\begin{equation}
H_{a}=-\frac{1}{2}\left( E_{el}\sigma _{z}+E_{J}\sigma _{x}\right) ,
\label{H}
\end{equation}%
where $E_{el}$ is the electrostatic energy and $E_{J}$ is the Josephson
coupling energy. The Pauli matrices $\sigma _{z}$ and $\sigma _{x}$ are in
the charge basis; that is, the basis states correspond to either zero or one
excess Cooper-pair charges on the island. These CPBs can be viewed as
artificial atoms with large dipole moments, and in circuit QED they are
coupled to microwave frequency photons in a quasi-one-dimensional
transmission line cavity (a coplanar waveguide resonator) by an electric
dipole interaction \cite{WallraffNAT04}. This apparatus has a number of
\textit{in situ} tunable parameters, including $E_{el}$ and $E_{J}$, and a
choice can be made \cite{BlaisPRA04} in such a way that the combined
Hamiltonian for qubit and transmission line cavity is the well-known
Jaynes-Cummings Hamiltonian
\begin{equation}
H_{JC}=\hbar \left( \omega _{r}a^{\dagger }a+\frac{1}{2}\right) +\frac{\hbar
\omega _{a}}{2}\sigma _{z}+\hbar \lambda \left( a^{\dagger }\sigma
_{-}+a\sigma _{+}\right) ,  \label{HJC}
\end{equation}%
where $a$ ($a^{\dagger }$) is the annihilation (creation) operator for the
cavity mode, $\omega _{r}$ is the cavity resonance frequency, $\omega _{a}$
is the energy splitting of the qubit, and $\lambda $ is the coupling
strength. Of course, the qubit energy splitting $\omega _{a}$ is a function
of the Cooper-pair box parameters $E_{el}$ and $E_{J}$ \cite{BlaisPRA04}.
For typical values of the parameters in this Hamiltonian see Refs. \cite%
{WallraffNAT04,BlaisPRA04}. It should be noted here that there has been a
basis change between Eqs. (\ref{H}) and (\ref{HJC}). In essence, we have
swapped the $x$ and $z$ axes of Eq. (\ref{H}), and the computational basis
for our qubit has become the Josephson basis of the CPB. This choice of
basis, which corresponds to operating the CPB in what is called the \textit{%
charge degeneracy point} $(E_{el}\sim 0)$, has a number of advantages: the
first of them being that the computational basis states become first-order
insensitive to dephasing from offset charge noise. In fact, a CPB is only
effective as a robust qubit at this operating point.

Single qubit gates are realized by pulses of microwaves on the input port of
the resonator. Depending on the frequency, phase, and amplitude of the
drive, different logical operations can be realized. External driving of the
resonator can be described by the Hamiltonian \cite{BlaisPRA07}%
\begin{equation}
H_{D}=\sum_{k}\left[ \epsilon _{k}(t)a^{\dagger }e^{-i\omega
_{d_{k}}}+\epsilon _{k}^{\ast }(t)ae^{i\omega _{d_{k}}}\right] ,
\end{equation}%
where $\epsilon _{k}(t)$ is the amplitude and $\omega _{d_{k}}$ the
frequency of the $k-$th external drive. In this scenario, quantum
fluctuations in the drive are very small with respect to the drive amplitude
and the drive can be considered, for all practical purposes, as a classical
field. In the case where the drive amplitude $\epsilon $ (single drive) is
independent of time, and by moving to a frame rotating at the frequency $%
\omega _{d}$ for both the qubit and the field operators, we get the
displaced Hamiltonian \cite{BlaisPRA07},
\begin{equation}
\widetilde{H}=\Delta _{r}a^{\dagger }a+\frac{\Delta _{a}}{2}\sigma
_{z}-g\left( a^{\dagger }\sigma _{-}+a\sigma _{+}\right) +\frac{\Omega _{R}}{%
2}\sigma _{x},  \label{HI}
\end{equation}%
In the foregoing equation $\Delta _{r}=\omega _{r}-\omega _{d}$\ stands for
the detuning of the cavity and the drive, $\Delta _{a}=\omega _{a}-\omega
_{d}$ is the same with respect with qubit transition frequency and the
drive, and $\Omega _{R}=2\epsilon g/\Delta _{r}$ is the Rabi frequency.
Changing $\Delta _{a}$, $\Omega _{R}$ and the phase of the drive can be used
to rotate the qubit around any axis on the Bloch sphere \cite{CollinPRL04}.
In the dispersive regime where $\Delta =\omega _{a}-\omega _{r}$ is much
bigger than the coupling $\lambda $, we can write the following aproximated
Hamiltonian \cite{BlaisPRA07}
\begin{equation}
H_{x}\approx \Delta _{r}a^{\dagger }a+\frac{\tilde{\Delta}_{a}}{2}\sigma
_{z}+\frac{\Omega _{R}}{2}\sigma _{x},
\end{equation}%
where we have defined $\chi =g^{2}/\Delta $ and $\tilde{\Delta}_{a}=\tilde{%
\omega}_{a}-\omega _{d}$ with $\tilde{\omega}_{a}=\omega _{a}+\chi $.

Rotations about the $z$ axis were produced from current pulses on the qubit
bias line that adiabatically change the qubit frequency, leading to phase
accumulation between the $\left\vert 0\right\rangle $ and $\left\vert
1\right\rangle $ states of the qubit \cite{SteffenPRL96}. Rotations about
any axis in the $xy$ plane were produced by microwave pulses resonant with
the qubit transition frequency. They selectively address only the qubit
energy levels, because transitions to higher-lying energy levels are
off-resonance due to anharmonicities of the potential and the shaping of the
pulses \cite{SteffenPRB03}. The phase of the microwave pulses defines the
rotation axis in the $xy$ plane. The \ rotation angle is controlled by the
pulse duration and amplitude.

In the situation where many qubits are fabricated with different transition
frequencies\ in the resonator, the qubits can be individually addressed by
tuning the frequency of the drive accordingly. It should therefore be
possible to individually control several qubits in the circuit QED
architecture. Assume the qubit $1$ initially in the superposition state
\begin{equation}
|\phi \rangle _{1}=C_{0}|\downarrow \rangle _{1}+C_{1}|\uparrow \rangle _{1},
\end{equation}%
where $C_{0}$ and $C_{1}$ are unknown coefficients. The resonator $R$ and
the qubit $2$ (receiver of teleported state) are prepared in the entangled
state $\left( |0\rangle _{R}|\uparrow \rangle _{2}-i|1\rangle
_{R}|\downarrow \rangle _{2}\right) /\sqrt{2}$. We assume one qubit coupled
with the resonator, the other uncoupled. Thus, the interaction described by
the Jaynes-Cummings Hamiltonian in Eq. (\ref{HJC}) creates the nonlocal
channel.

The state for the whole system can be written in the form%
\begin{eqnarray}
|\psi \rangle &=&\frac{1}{2}\left[ |\Psi ^{(+)}\rangle _{1R}\left(
C_{0}|\downarrow \rangle _{2}+C_{1}|\uparrow \rangle _{2}\right) \right.
\notag \\
&+&\left. |\Psi ^{(-)}\rangle _{1R}\left( C_{0}|\downarrow \rangle
_{2}-C_{1}|\uparrow \rangle _{2}\right) \right.  \notag \\
&+&\left. |\Phi ^{(+)}\rangle _{1R}\left( C_{0}|\uparrow \rangle
_{2}+C_{1}|\downarrow \rangle _{2}\right) \right.  \notag \\
&+&\left. |\Phi ^{(-)}\rangle _{1R}\left( C_{0}|\uparrow \rangle
_{2}-C_{1}|\downarrow \rangle _{2}\right) \right]
\end{eqnarray}%
where $|\Psi ^{(\pm )}\rangle _{1R}$ and $|\Phi ^{(\pm )}\rangle _{1R}$ are
the Bell states \cite{BraunsteinPRL92}
\begin{equation}
|\Psi ^{(\pm )}\rangle _{1R}=\frac{1}{\sqrt{2}}\left( -i|\downarrow \rangle
_{1}|1\rangle _{R}\pm |\uparrow \rangle _{1}|0\rangle _{R}\right) ,
\end{equation}%
\begin{equation}
|\Phi ^{(\pm )}\rangle _{1R}=\frac{1}{\sqrt{2}}\left( |\downarrow \rangle
_{1}|0\rangle _{R}\pm i|\uparrow \rangle _{1}|1\rangle _{R}\right) .
\end{equation}

The outcome of the joint measurement on the qubit $1$ and the resonator $R$
is transmitted to the receiver, who can apply an appropriate rotation to
qubit $2$ to reconstruct the initial state of qubit $1$. The measurement of
the entanglement of the qubit-resonator system can be done through state
tomography \cite{SteffenSC06}. The multiqubit states teleportation, such as
GHZ, W, and other entangled qubits, can also be implemented in
superconducting circuits in the same way proposed in this paper.

In conclusion, we proposed a theoretical scheme to implement a quantum
teleportation based on the current experimental design \cite%
{WallraffNAT04,SchusterNAT07} in which superconducting charge qubits are
capacitively coupled to a single high-Q superconducting coplanar resonator.
In this system the resonator frequency $\omega _{r}/2\pi $ is assumed in the
interval $5$ - $10$ GHz. The qubit transition frequencies were chosen as
belonging to the interval $5$ - $15$ GHz, tunable via a flux though the
qubit loop. In the circuit both qubits are affected by the externally
applied field, but the effect on each qubit depends on the area of the
qubit's loop. Coupling strengths $g/2\pi $ between $5.8$ and $100$ MHz have
been realized experimentally and couplings up to $200$ MHz should be
feasible \cite{WallraffNAT04,SchusterNAT07}. Rabi frequencies of $50$ MHz
were obtained with a sample of moderate coupling strength $g/2\pi =17$ MHz\
and an improvement by at least a factor of 2 is realistic \cite%
{WallraffPRL95}. The cavity damping rate is chosen at fabrication time by
tuning the coupling capacitance between the resonator center line and its
input and output ports. Quality factors up to $Q\sim 10^{6}$ have been
reported for undercoupled resonators \cite{FrunzioIEEE05,DayNAT03},
corresponding to a low damping rate $\kappa /2\pi =\omega _{r}/2Q\sim 5$ KHz
for a $\omega _{r}/2\pi =5$ GHz resonator. This results in a long photon
lifetime $1/\kappa $ of $31$ $\mu s$. To allow for fast measurement, the
coupled quality factor can also be reduced by two or more orders of
magnitude. Relaxation and dephasing of a qubit of this system were measured
in Ref. \cite{WallraffPRL95}, where $T_{1}=7.3$ $\mu s$ and $T_{2}=500$ $ns$
were reported. These yield $\gamma _{1}/2\pi \ =0.02$ MHz and \ $\gamma
_{\phi }/2\pi =\left( \gamma _{2}-\gamma _{1}/2\right) /2\pi =0.31$ MHz. In
the limit where $\Delta _{r}$ is large compared with the resonator
half-width $\kappa /2$, the average photon number in the resonator can be
written as $\overline{n}\approx (\epsilon /\Delta _{r})^{2}$. In this case,
the Rabi frequency takes the simple form $\Omega _{R}\approx 2g\sqrt{%
\overline{n}}$ , as expected from the Jaynes-Cummings model.

The Cooper pair box is especially well suited for cavity QED because of its
large effective electric dipole moment $d$, which can be $10^{4}$ times
larger than in alkali atoms and ten times larger in typical Rydberg atoms
\cite{WallraffNAT04}. Besides the advantage of implementing teleportation
with the present devices, in comparisom with the cavity QED, coming from the
easily reproducible architeture, the efficiency, and low loss, another great
advantage of using CPBs comes from the velocity of the teleportation
proccess, very important for quantum computation and information, in view,
e.g., of the unavoidable presence of decoherence effects.

We thank the CAPES (RMG), CNPq (BB), and Funape (WBC, ATA, BB), Brazilian
Agencies, for partially supporting this work.

\end{document}